\documentclass{article}
\makeatletter \@addtoreset{equation}{section} \makeatother
\renewcommand{\theequation}{\arabic{section}$.$\arabic{equation}}

\begin{document}
\date{\today}

\title{\bf On the finite-size behavior of systems \\
with asymptotically large critical shift}
\author{Daniel M. Dantchev$^{1,2,3}$\thanks{e-mail:
daniel@imbm.bas.bg} \ and Jordan G. Brankov$^{1}$\thanks{e-mail:
brankov@imbm.bas.bg}\\ \\
$^1$Institute of Mechanics, Bulgarian
Academy of Sciences, \\Acad.
G. Bonchev St. bl. 4, 1113 Sofia, Bulgaria,\\
$^2$Max-Planck-Institut f\"{u}r Metallforschung,
\\Heisenbergstrasse 3, D-70569 Stuttgart, Germany
\\
$^3$Institut f\"{u}r Theoretische und Angewandte Physik,\\
Universit\"{a}t Stuttgart, Pfaffenwaldring 57, D-70569 Stuttgart,
Germany}

\maketitle
\begin{abstract}
Exact results of the finite-size behavior of the susceptibility in
three-dimensional mean spherical model films under
Dirichlet-Dirichlet, Dirichlet-Neumann and Neumann-Neumann
boundary conditions are presented. The corresponding scaling
functions are explicitly derived and their asymptotics close to,
above and below the bulk critical temperature $T_c$ are obtained.
The results can be incorporated in the framework of the
finite-size scaling theory where the exponent $\lambda$
characterizing the shift of the finite-size critical temperature
with respect to $T_c$ is smaller than $1/\nu$, with $\nu$ being
the critical exponent of the bulk correlation length.
\end{abstract}


\section{Introduction}
The basic ideas of the phenomenological finite-size scaling theory
at criticality have been suggested by Fisher \cite{F71}, and Fisher and
Barber \cite{FB72} (for more recent reviews consult \cite{Ba83},
\cite{P90}, and \cite{BDT00}). According to the phenomenological
theory, rounding and shifting of the anomalies in the
thermodynamic functions set in when the bulk correlation length
$\xi_{\infty}$ becomes comparable to the characteristic linear
size $L$ of the system. More specifically, it is predicted that
finite-size effects are controlled by the ratio $L/\xi_{\infty}$.
Here we recall  some fundamental notions and facts of that theory.

Let us start with a system having the geometry
$L^{d-d'}\times\infty^{d'}$, with  $d'>d_l$, where $d_l$ is the
lower critical dimension of the corresponding class of bulk
systems. Then, under boundary conditions $\tau$ imposed across the
finite dimensions of the system, the {\it finite-size} system
exhibits a phase transition at a temperature $T=T_{c,L}^{(\tau)}$,
and the corresponding infinite system at
$T=T_{c,\infty}^{(\tau)}\equiv T_c$. The so-called {\it fractional
shift}, characterizing the shift of the critical temperature of
the finite-size system\index{critical temperature!fractional
shift} is defined as
\begin{equation}\label{fracshift}
\varepsilon_L^{(\tau)} =
\left(T_c-T_{c,L}^{(\tau)}\right)/T_c\simeq b^{(\tau)}
L^{-\lambda},
\end{equation}
where the expected asymptotic behavior for $L\gg 1$ is given by the
shift exponent $\lambda$.

Let us now consider  the susceptibility $\chi$ (per spin) which in
the bulk (infinite) system has a critical-point divergence of the
type
\begin{equation}\label{Ybulk}
\chi_{\infty}(T)\simeq A t^{-\gamma},\qquad t\rightarrow 0^+,
\end{equation}
where $\gamma=\gamma(d)$ is the $d$-dimensional critical exponent
and $t=(T-T_c)/T_c$. On approaching the finite-size critical
temperature from above at fixed $L$ one should have
\begin{equation}\label{Yfinite}
\chi_L^{(\tau)}(T)\simeq \dot{A}_L^{(\tau)}
\dot{t}^{-\dot{\gamma}}, \qquad  \dot{t}\rightarrow 0^+,
\end{equation}
where
$
\dot{t}=\left(T-T_{c,L}^{(\tau)}\right)/T_c =
\varepsilon_L^{(\tau)}+t
$
and $\dot{\gamma}=\gamma (d')$ (in general, $\gamma \neq
\dot{\gamma}$). Let $T_{*,L}^{(\tau)}$ denote the temperature at
which the considered finite-size property $\chi_L^{(\tau)}(T)$
first shows significant (of the relative order of unity) deviation
from its bulk limit $\chi_{\infty}(T)$. Then one defines the {\it
fractional rounding} \cite{FB72} $
\delta_L^{(\tau)}=\left(T_{*,L}^{(\tau)}-T_c \right)/T_c \simeq
c^{(\tau)}L^{-\theta}, \ L\gg 1.
$ The ``rounding'' measures the region of crossover from bulk
$d$-dimensional to $d'$-dimensional critical behavior.

The basic assertions of the "orthodox" {\it phenomenological
finite-size scaling} are that {\it (i)} the only relevant variable
on which the properties of the finite-size system depend in the
neighborhood of $T_c$ is $L/\xi_{\infty}(T) \sim Lt^\nu$, and {\it
(ii)} the rounding occurs when $\xi_{\infty}(T)\simeq L$.

It is easy to see that assumption {\it (ii)} leads directly to the
conclusion that $\theta=1/\nu$ and, from {\it (i)}, it immediately
follows that
\begin{equation}\label{basicscaling}
\chi_L^{(\tau)}(T)\simeq L^{\gamma/\nu}
\tilde{X}^{(\tau)}\left(L/\xi_{\infty}(T)\right),
\end{equation}
or, equivalently,
\begin{equation}\label{Yscaling}
 \chi_L^{(\tau)}(T)\simeq L^{\gamma/\nu} X^{(\tau)}(tL^{1/\nu}).
\end{equation}
Here $X^{(\tau)}(x)$ is the universal {\it finite-size scaling
function}\index{finite-size scaling!function} describing the
critical behavior of $\chi$, where, in order to reproduce the
behavior described by Eqs. (\ref{Ybulk}) and (\ref{Yfinite}) one
should have:
\begin{equation}\label{Xasp}
 X^{(\tau)}(x)\simeq X_\infty x^{-\gamma},\quad \mbox{as}
\quad  x\rightarrow \infty,
\end{equation}
and
\begin{equation}\label{tdotbehavior}
X^{(\tau)}(x)\simeq X_0^{(\tau)} x^{-\dot{\gamma}} \quad \mbox{as}
\quad x\rightarrow 0^+.
\end{equation}

It has been considered, see \cite{F71,FB72,Ba83}, that a more
general formulation of the finite-size scaling hypothesis is given
by the equation
\begin{equation}\label{Yscalingmod}
 \chi_L^{(\tau)}(T)\simeq L^{\gamma/\nu} X^{(\tau)}(\dot{t}L^{1/\nu}).
\end{equation}
Apart from the allowed shift of $T_c$ from $T_{c,L}$, Eqs.
(\ref{Yscaling}) and (\ref{Yscalingmod}) are equivalent if
$\xi_{\infty}$ diverges algebraically with exponent $\nu \geq
1/\lambda$.

We emphasize that the use of the shifted temperature variable
$\dot{t}$ in the above finite-size scaling hypotheses allows for
any $L$-dependence of the shift $\varepsilon^{(\tau)}_L$, i.e. the
shift exponent $\lambda$ remains arbitrary. The assertion that the
only criterion determining the finite-size scaling effects in the
critical region is $\xi_{\infty}(T) \simeq L$ leads to the
equalities $ \lambda=\theta=1/\nu$. This result follows from the
renormalization group derivation of finite-size scaling
\cite{Ba83}, \cite{B82}, see also \cite{BM78}. Except in some
special cases (ideal Bose gas and spherical model with a film
geometry and Dirichlet-Dirichlet \cite{BF73}, \cite{B74},
\cite{D93}, \cite{CD} or Neumann-Neumann boundaries \cite{DBA97a},
when  one has a logarithmic shift of the type $\pm \ln L/L$ for
$d=3$ and $\lambda=1$ in {\it all} other dimensions $d>2$), this
relation seems to be quite generally valid. We stress,
nevertheless, that the relationship $\lambda=1/\nu$ is {\it not}
\cite{FB72} a necessary condition for the finite-size scaling to
hold in general.

\subsection*{Phenomenological finite-size scaling for systems with large critical shift}
Let us now consider in a bit more details what will be the
consequences if the shift is asymptotically large, i.e. when
$1/\nu>\lambda$. From Eq. (\ref{Yscalingmod}) one then has
\begin{equation}\label{theothercase}
 \chi_L^{(\tau)}(T)\simeq L^{\gamma/\nu} X^{(\tau)}(t L^{1/\nu}+
 b^{(\tau)}L^{1/\nu-\lambda}).
\end{equation}
Obviously, in order to make explicit statements, we have to
consider the two possibilities {\it i)} $b^{(\tau)}>0$ and {\it
ii)} $b^{(\tau)}<0$ separately.

\subsubsection*{Case {\it i)} : $b^{(\tau)}>0$}

From Eq. (\ref{Xasp}) it immediately follows that at $T_c$
\begin{equation}\label{plus}
    \chi_L^{(\tau)}(T_c)\simeq X_\infty
    \left[b^{(\tau)}\right]^{-\gamma}L^{\gamma \lambda},
\end{equation}
i.e., the divergence of the susceptibility at $T_c$ with respect to
$L$ will be reduced in comparison with the "standard" behavior
\begin{equation}\label{atTC}
\chi_L^{(\tau)}(T_c)\simeq  X^{(\tau)}(0) L^{\gamma/\nu}
\end{equation}
predicted by Eq. (\ref{Yscaling}).

An example of a model with large positive shift of the critical
temperature is the spherical model under Neumann-Neumann boundary
conditions \cite{DBA97a} (see below). For $d=3$ the shift in the
dimensionless critical coupling is equal
to $\ln L/(4\pi L)$. For such a shift one immediately obtains that
\begin{equation}\label{plussm}
    \chi_L^{(\tau)}(T_c)\simeq X_\infty
    \left[\frac{4\pi L}{\ln L}\right]^{\gamma}.
\end{equation}

\subsubsection*{Case {\it ii)} : $b^{(\tau)}<0$}

 Then, at $T_c$ one has $\dot{t}L^{1/\nu}=b^{(\tau)}L^{1/\nu-\lambda}
\rightarrow  -\infty$ when
$L\rightarrow\infty$. Obviously, in order to give a general answer
what will be the behavior of the susceptibility in this case, one
needs to know the asymptotics of the scaling function
$X^{(\tau)}(x)$ for $x\rightarrow -\infty$.

Generally speaking, when $x=tL^{1/\nu}\rightarrow -\infty$ takes
place, the behavior of the zero-field susceptibility in a $O(n)$
model depends on the fact whether $n=1$, or $n>1$, and on the geometry
of the system. Then, under periodic boundary conditions,
summarizing the results of \cite{PF83}, \cite{P84}, \cite{BZ85}
for the Ising type models and that ones of \cite{B82},
\cite{FP85}, \cite{SP85}, \cite{BZ85}, \cite{SP86}, \cite{BD91}
for $O(n)$, $n\ge 2$ models, one has
\begin{equation}\label{Ybelow}
\chi_L^{(p)}(T < T_c) \sim \left\{
\begin{array}{ll}
L^d, &d'=0\\[1mm]
L \exp[L^{d-1}\sigma(T)/k_B T], &d'=1, \qquad n=1\\[1mm]
L^{2(d-1)}, &d'=1, \qquad n \ge 2 \\[1mm]
L^2 \exp[c L^{d-2}\Upsilon(T)/k_B T], &d'=2,\qquad n\ge 2\\[1mm]
\infty, & \!\!\!\begin{array}{ll} \mbox{if} \ d'>1 \ \mbox{and} &
n=1,\\[1mm]
                          \mbox{or} \ d'>2 \ \mbox{and} & n \ge 2,
         \end{array}
\end{array}
 \right.
\end{equation}
where $\sigma(T)$ is the interfacial (or surface) tension in the
Ising model, $\Upsilon(T)$ is its analog - the helicity modulus -
for an $O(n)$, $n \ge 2$ system, and $c$ is a constant. For other
than periodic boundary conditions the situation is
not so clear and a detailed information for the behavior of the
zero-field finite-size susceptibility is still to be established,
but one can hope that the leading-order behavior will remain unchanged
(eventually, the power of $L$ in front of the exponential terms may
change when $d'=d_l$).

We shall demonstrate the consequences of the shift on the finite-size
behavior of the susceptibility for a fully finite and a
film geometry only. The interested reader can easily
complete the list of all possible geometries.

\subsubsection*{{\it The fully finite geometry}}

In order to reproduce the size dependence of the susceptibility
at $T<T_c$, one should have $X^{(\tau)}(x)\simeq
X_-^{(\tau)}x^{d\nu-\gamma}$. Then at $T_c$ one obtains
\begin{equation}\label{minus}
    \chi_L^{(\tau)}(T_c)\simeq X_-
    \left|b^{(\tau)}\right|^{d\nu-\gamma}L^{d-\lambda\nu(d-2+\eta)}.
\end{equation}
For the spherical model under Dirichlet boundary conditions,
taking into account that $\lambda=1$, $\eta=0$ and $\nu=1/(d-2)$
(for $2< d< 4$) one derives
\begin{equation}\label{minus2}
    \chi_L^{(\tau)}(T_c)\simeq A L^{d-1},
\end{equation}
where $A$ is a constant. This is exactly the result recently
reported in \cite{CD}.

\subsubsection*{{\it The film geometry}}

Similarly to the above, one can write down again the general
expressions, but since they are rather cumbersome, we shall
consider only the case of a film geometry on the example of the
three-dimensional spherical model. Taking into account that then
$\beta\Upsilon(T)=K-K_c$ \cite{D93}, \cite{FBJ73}, where $K$ is
the dimensionless coupling in the system, and that
$\varepsilon_L^{(D)}=-\ln L/(4\pi L)$, $c=4\pi$ \cite{BF73},
\cite{B74}, \cite{D93}, \cite{CD}, we obtain
\begin{equation}\label{minusfilm}
    \chi_L^{(\tau)}(T_c)\sim L^{2}\exp(c \ln L/4\pi)\sim L^3,
\end{equation}
which is exactly the result recently reported in \cite{CD} (see
also below). Note that now $\chi_L^{(\tau)}$ diverges at $T_c$
faster than in the case of the "standard" finite-size scaling
prediction given by Eq. (\ref{atTC}) (we recall that
$\gamma=2/(d-2)$ for the spherical model, i.e. $\gamma/\nu=2$). It
is clear that when $d'=2$, but $d > 3$, one will have
$\chi_L^{(\tau)}(T_c)\sim L^{2}\exp(\tilde{c} L^{d-3})$, where
$\tilde{c}$ is a constant, and we have taken into account that
$\lambda=1$ for all $d>2$.

Before finishing this overview of the general form of the
phenomenological finite-size scaling theory and its consequences,
let us just remind the reader what the finite-size behavior in the
case of geometry $L^{d-d'}\times \infty^{d'}$ with $d'\le d_l$ is.
Then, in the finite-size system there is no phase transition of
its own, therefore, $\dot{\gamma} \equiv 0$. The hypotheses stated
above still hold, provided we set $\dot{\gamma}= 0$ and replace
$T_{c,L}^{(\tau)}$ by appropriately defined {\it pseudocritical}
temperature $T^{(\tau)}_{m,L}$. The latter can be defined, for
example, as the temperature at which the susceptibility reaches
its maximum. Then, in the case of an algebraic bulk singularity of
the type (\ref{Ybulk}), one has, e.g., $
\chi_L^{(\tau)}(T^{(\tau)}_{m,L}) \simeq
X_0^{(\tau)}L^{\gamma/\nu}, \ L\gg 1.
$
This asymptotic behavior is exploited in one of the basic methods
for evaluation of bulk critical exponents from finite-size data.
Here we would like to stress that $T^{(\tau)}_{m,L}$ depends
on which physical quantity has been used for its definition, since
the maxima of the susceptibility and, say, the specific heat
take place at, generally speaking, different temperatures.

In the next sections we will present exact results
for the finite-size behavior of the zero-field susceptibility of
the mean spherical model with a film geometry when
Dirichlet-Dirichlet, Neumann-Dirichlet and
Neumann-Neumann boundary conditions are applied at the surfaces
of the film. The corresponding scaling functions
$X^{(\tau)}$ will be derived and their asymptotic behavior will  be
analyzed. We will demonstrate the important role of the shift when
it is asymptotically larger than $L^{-1/\nu}$.

The structure of the article is as follows. In Section \ref{model}
we briefly present the model and the basic facts needed for the
analytical treatment of the free energy and the susceptibility
for a fully finite system. Section \ref{film} contains the corresponding
modifications of these expressions in the case of a film geometry.
The finite-size critical behavior  of the susceptibility is
analyzed in Section \ref{finite_size}. The article closes with a
discussion of the results obtained and their eventual
generalization. Some technical details needed for evaluation of
different sums are described in the Appendix.

\section{The model}
\label{model}

We consider a $d$-dimensional mean spherical model with
nearest-neighbor ferromagnetic interactions on a  simple cubic
lattice. At each lattice site $\vec{r} =
(r_{1},r_{2},\cdots,r_{d}) \in {\bf Z}^{d}$ there is a random
(spin) variable $\sigma(\vec{r}) \in {\bf R}$ and
$\sigma_{_{\Lambda}} = \{ \sigma(\vec{r}), \vec{r}\in \Lambda \}$
is the configuration in a finite region $\Lambda \subset {\bf
Z}^{d}$, containing $|\Lambda|$ sites. The boundary conditions (to
be denoted by the superscript $\tau$) define the interaction of
the spins in the region $\Lambda$ with a specified configuration
$\{\sigma(\vec{r}), \vec{r} \in \Lambda^{c} \}$ in the complement
$\Lambda^{c} = {\bf Z}^{d} \setminus \Lambda$. In the remainder we
take $\Lambda$ to be the parallelepiped $\Lambda = {\cal L}_{1}
\times {\cal L}_{2} \times \cdots \times {\cal L}_{d}$, with
${\cal L}_{i} = \{1,\dots ,L_{i}\}$, and explicitly study the case
of film geometry which results in the limit $L_{2},\cdots,L_{d}
\rightarrow \infty$ at finite values of $L_{1} = L$. In the finite
$r_{1}$ direction it suffices to specify the values of
$\sigma(0,r_{2},\cdots,r_{d})$ and $\sigma(L+1,r_{2},\cdots,
r_{d})$ for all $(r_{2},\cdots,r_{d}) \in {\cal L}_{2} \times
\cdots \times {\cal L}_{d}$.

The finite-size scaling behavior of the mean spherical model has
been studied so far under periodic, antiperiodic, Dirichlet,
\cite{BF73}, \cite{B74}, \cite{BJSW}, \cite{SJB}, Neumann
\cite{DBA97a}, \cite{DBA97b} and Neumann-Dirichlet \cite{DBA97a}
boundary conditions (for a review see, e.g., \cite{BDT00}). For
lattice systems under Dirichlet boundary conditions we mean that
\begin{equation}
\sigma(0,r_{2},\cdots,r_{d})=\sigma(L+1,r_{2},\cdots,r_{d})=0,\label{Dirichlet}
\end{equation}
under Neumann boundary conditions that
\begin{equation}\label{Neumann}
  \sigma(0,r_{2}, \cdots,
r_{d})= \sigma(1,r_{2},\cdots, r_{d}), \ \ \ \sigma(L+1,r_{2},
\cdots, r_{d})= \sigma(L,r_{2}, \cdots,r_{d}),
\end{equation}
and, under Neumann-Dirichlet boundary conditions that
\begin{equation}\label{ND}
\sigma(0,r_{2},\cdots, r_{d})= \sigma(1,r_{2},\cdots, r_{d}), \ \
\ \sigma(L+1,r_{2},\cdots, r_{d})= 0.
\end{equation}
Obviously, the terminology Dirichlet and Neumann boundary
conditions is justified by analogy with the continuum limit. The
case of free surfaces in a system of film geometry (in the limit
$L_{2}, \cdots, L_{d} \rightarrow \infty$), considered in the
literature \cite{BF73}, \cite{B74}, \cite{BJSW}, \cite{SJB},
corresponds to Dirichlet boundary conditions ($\tau\equiv D$).
When a fully finite system is envisaged, we always assume periodic
boundary conditions with respect to the coordinates $r_2,\cdots
,r_d$ , i.e., for all $\vec{r} \in \Lambda$ and all integers
$m_2,\cdots,m_d$, we set $\sigma(r_{1},r_{2}+m_2L_{2},\cdots,
r_{d}+m_dL_{d})=\sigma(r_{1},r_{2}, \cdots, r_{d}) $.

For brevity of notation, we consider the configuration space
$\Omega_{_{\Lambda}} = {\bf R}^{|\Lambda|}$ as an Euclidean vector
space in which each configuration is represented by a
column-vector $\sigma_{_{\Lambda}}$ with components labelled
according to the lexicographic order of the set $\{ (r_{1},r_{2},
\cdots, r_{d}) \in \Lambda \}$. Let $\sigma_{_{\Lambda}}^{\dag}$
be the corresponding transposed row-vector and let the dot
($\cdot$) denote matrix multiplication. Then, for given boundary
conditions $\tau = (\tau_{1}, \tau_{2}, \cdots, \tau_{d})$,
specified for each pair of opposite faces of $\Lambda$ by some
$\tau_{i} = p$ (periodic), $D$ (Dirichlet), $N$ (Neumann), or $ND$
(Neumann-Dirichlet), and given external magnetic field configuration
$h_{_{\Lambda}} = \{h(\vec{r}), \vec{r} \in \Lambda \}$, with $h(\vec{r})
\in {\bf R}$, the Hamiltonian of the model takes the form
\begin{equation}
\beta {\cal H}^{(\tau)}_{{\Lambda}}(\sigma_{{\Lambda}}|K,
h_{{\Lambda}};s) = -\frac{1}{2} K \sigma_{{\Lambda}}^{\dag} \cdot
Q_{{\Lambda}}^{(\tau)} \cdot \sigma_{{\Lambda}} + s\:
\sigma_{{\Lambda}}^{\dag} \cdot \sigma_{_{\Lambda}} -
h_{{\Lambda}}^{\dag} \cdot \sigma_{{\Lambda}}.
\label{hamdiagonal}
\end{equation}
Here $\beta = 1/k_{B}T$ is the inverse temperature; $K=\beta J$ is
the dimensionless coupling constant; $s$ is the spherical field
which is to be determined from the mean spherical constraint, see
equation (\ref{constraint}) below; the $|\Lambda| \times |\Lambda|$
interaction matrix $Q^{(\tau)} _{{\Lambda}}$ can be written as
\begin{equation}
Q^{(\tau)}_{{\Lambda}} = (\Delta_{1}^{(\tau_{1})}+2\: E_{1})
\times \cdots \times (\Delta_{d}^{(\tau_{d})}+2\: E_{d}),
\label{matrix}
\end{equation}
where $\times$ denotes the outer product of the corresponding
matrices, $\Delta_{i}^{(\tau_{i})}$ is the $L_{i} \times L_{i}$
discrete Laplacian under boundary conditions $\tau_{i}$, and
$E_{i}$ is the $L_{i} \times L_{i}$ unit matrix.

As it is well known, the complete set of orthonormal
eigenfunctions, $\{u_{_{L}}^{(\tau)}(r,k)$, $k=1,\dots ,L\}$, of
the one-dimensional discrete Laplacian for periodic, Dirichlet,
Neumann and Neumann-Dirichlet boundary conditions is given by
\begin{equation}
u_{_{L}}^{(p)}(r,k) =L^{-1/2} \exp \left[-{\mathrm i} \:
r\varphi_{_{L}}^{(p)}(k)\right], \label{eigenvecP}
\end{equation}
\begin{equation}
u_{_{L}}^{(D)}(r,k) = [2/(L+1)]^{1/2}\sin \left[
r\varphi_{_{L}}^{(D)}(k)\right], \label{eigenvecD}
\end{equation}
\begin{equation}
u_{_{L}}^{(N)}(r,k) =
\left\{
\begin{array}{ll} L^{-1/2} & \mbox{for} \ \ k=1  \\  (2/L)^{1/2} \cos
\left[ (r-\frac{1}{2})\varphi_{_{L}}^{(N)}(k)\right] & \mbox{for} \
\ k=2,\cdots,L
\end{array} \right.
\label{eigenvecN}
\end{equation}
and
\begin{equation}
u_{_{L}}^{(ND)}(r,k) = 2(2L+1)^{-1/2}\cos \left[
\left(r-\frac{1}{2}\right)\varphi_{_{L}}^{(ND)}(k)\right],
\label{eigenvecND}
\end{equation}
where
\begin{eqnarray}
&& \varphi_{_{L}}^{(p)}(k) = {2\pi k\over L}, \ \ \
\varphi_{_{L}}^{(D)}(k) = {\pi k\over L+1}, \\
&& \varphi_{_{L}}^{(N)}(k) = {\pi (k-1) \over L}, \ \ \
\varphi_{_{L}}^{(ND)}(k) = {\pi (2k-1) \over 2L+1}. \label{eigenv}
\end{eqnarray}
The corresponding eigenvalues are
\begin{equation}
\lambda_{_{L}}^{(\tau)}(k) = -2+2\cos \varphi_{_{L}}^{(\tau)}(k),
\hspace{1cm} k= 1,\dots ,L. \label{eigenvalues1}
\end{equation}
The eigenfunctions of the interaction matrix (\ref{matrix}) have
the form
\begin{equation}
u_{_{\Lambda}}^{(\tau)}(\vec{r},\vec{k}) =
u_{_{L_{1}}}^{(\tau_{1})} (r_{1},k_{1}) \:
u_{_{L_{2}}}^{(\tau_{2})}(r_{2},k_{2}) \: \cdots \:
u_{_{L_{d}}}^{(\tau_{d})}(r_{d},k_{d}),\hspace{1cm} \vec{k}\in
\Lambda , \label{eigenfunctions}
\end{equation}
and the corresponding eigenvalues are
\begin{equation}
\mu_{_{\Lambda}}^{(\tau)}(\vec{k}) = 2\sum_{\nu = 1}^{d}\cos
\varphi_{_{L_{\nu}}}^{(\tau_{\nu})}(k_{\nu}), \hspace{1cm}
\vec{k}\in \Lambda . \label{eigenvalues}
\end{equation}

In order to ensure positivity of all the eigenvalues $-\frac{1}{2}
K \mu_{_{\Lambda}}^{ (\tau)}(\vec{k})+s$, $\vec{k} \in \Lambda $,
of the quadratic form in $\beta {\cal H}^{(\tau)}_{_{\Lambda}}
(\sigma_{_{\Lambda}}|K,h_{_{\Lambda}};s)$, see equation (\ref{hamdiagonal}),
it is convenient to introduce a shifted
and rescaled spherical field $\phi > 0$ by setting
\begin{equation}
s= s(\phi) :=\frac{1}{2} K[\phi +
\mu_{_{\Lambda}}^{(\tau)}(\vec{k}_{0})], \label{spfielddefinition}
\end{equation}
where $\vec{k}_{0}$ is a vector $\vec{k} \in \Lambda$ at which
$\mu_{_{\Lambda}}^{(\tau)}(\vec{k})$ attains maximum value.

The joint probability distribution of the random variables
$\sigma_{_{\Lambda}} = \{\sigma(\vec{r}), \vec{r} \in \Lambda \}$
is given by the Gibbs measure
\begin{equation}
d\rho_{{\Lambda}}^{(\tau)}(\sigma_{{\Lambda}}|K,h_{_{\Lambda}};\phi)=
\exp\left[-\beta {\cal
H}^{(\tau)}_{_{\Lambda}}(\sigma_{_{\Lambda}}|K,
h_{{\Lambda}};s(\phi ))\right]\: \prod_{\vec{r}\in \Lambda} d
\sigma (\vec{r})/ Z_{{\Lambda}}^{(\tau)}(K,h_{{\Lambda}};\phi),
\label{probmeasure}
\end{equation}
where $d \sigma (\vec{r})$ is the Lebesgue measure on ${\bf R}$
and
\begin{equation}
Z_{{\Lambda}}^{(\tau)}(K,h_{{\Lambda}};\phi) = \int_{{\bf
R}^{|\Lambda|}}\exp\left[- \beta {\cal H}^{(\tau)}_{{
\Lambda}}(\sigma_{{\Lambda}}|K,h_{_{\Lambda}};s(\phi))\right]\:
\prod_{\vec{r}\in \Lambda} d \sigma (\vec{r})
\label{partitionfunction}
\end{equation}
is the partition function of the Gaussian model. The latter
is finite for all $\phi > 0$ and equals $+\infty$ for $\phi \leq 0$.
The free-energy density of the  mean spherical model in a
finite region $\Lambda$ is given by the Legendre transformation
\begin{equation}
\beta f_{_{\Lambda}}^{(\tau)}(K,h_{_{\Lambda}}) := \sup_{\phi}
\left\{ -|\Lambda|^{-1}\ln
Z_{_{\Lambda}}^{(\tau)}(K,h_{_{\Lambda}};\phi) - s(\phi)\right\}.
\label{freeenergy}
\end{equation}
Here the supremum is attained at the solution $\phi =
\phi_{{\Lambda}}^{ (\tau)}(K,h_{{\Lambda}})$ (for brevity
denoted by $\phi_{{\Lambda}}^{(\tau)}$) of the mean spherical
constraint
\begin{equation}
|\Lambda|^{-1}\sum_{\vec{r}\in \Lambda}\langle \sigma^{2}
(\vec{r}) \rangle_{{\Lambda}}^{(\tau)}(K,h_{{\Lambda}};\phi) = 1,
\label{constraint}
\end{equation}
where $\langle \cdots
\rangle_{{\Lambda}}^{(\tau)}(K,h_{{\Lambda}};\phi)$ denotes
expectation value with respect to the measure (\ref{probmeasure}).

By direct evaluation of the integrals in the partition function
(\ref{partitionfunction}), one obtains
\begin{eqnarray}
\beta f_{{\Lambda}}^{(\tau)}(K,h_{{\Lambda}}) = \frac12 \left\{
\ln(K/2\pi) -
K\mu_{{\Lambda}}^{(\tau)}(\vec{k}_{0}) \right. \nonumber \\
\left. + U_{{\Lambda}}^{(\tau)}(\phi_{{\Lambda}}^{(\tau)}) -
P_{{\Lambda}}^{(\tau)}(K,h_{{\Lambda}}; \phi_{{\Lambda}}^{(\tau)})
- K\phi_{{\Lambda}}^{(\tau)}\right\}. \label{FE}
\end{eqnarray}
Here we have introduced the function
\begin{equation}
U_{_{\Lambda}}^{(\tau)}(\phi) = |\Lambda|^{-1}\sum_{\vec{k}\in
\Lambda} \ln [\phi + \omega_{_{\Lambda}}^{(\tau)}(\vec{k})],
\label{energy}
\end{equation}
which describes the contribution of the spin-spin interaction (called
"interaction term"), where
\begin{equation}
\omega_{{\Lambda}}^{(\tau)}(\vec{k}) :=
\mu_{{\Lambda}}^{(\tau)}(\vec{k}_{0})-\mu_{{\Lambda}}^{(\tau)}(\vec{k})
\end{equation}
is the normalized excitation spectrum,
 and the function
\begin{equation}
P_{{\Lambda}}^{(\tau)}(K,h_{_{\Lambda}};\phi) =
\frac{1}{K|\Lambda|} \sum_{\vec{k}\in \Lambda}
{|\hat{h}_{{\Lambda}}^{(\tau)} (\vec{k})|^{2} \over \phi +
\omega_{{\Lambda}}^{(\tau)}(\vec{k})}, \label{fieldterm}
\end{equation}
represents the "field term". In (\ref{fieldterm})
$\hat{h}_{{\Lambda}}^{(\tau)}(\vec{k})$ denotes the projection of
the magnetic field configuration $h_{{\Lambda}}$ on the
eigenfunction $\bar{u}_{{\Lambda}}^{(\tau)}(\vec{r},\vec{k})$:
\begin{equation}
\hat{h}_{{\Lambda}}^{(\tau)}(\vec{k}) = \sum_{\vec{r}\in \Lambda}
h(\vec{r}) \bar{u}_{{\Lambda}}^{(\tau)}(\vec{r},\vec{k}).
\label{thefield}
\end{equation}

The mean spherical constraint (\ref{constraint}) has the form
\begin{equation}
\frac{d}{d \phi}U_{{\Lambda}}^{(\tau)}(\phi)-{\partial \over
\partial \phi} P_{{\Lambda}}^{(\tau)}(K,h_{_{\Lambda}};\phi) =
K. \label{msconstraint}
\end{equation}
Its solution $\phi=\phi_{{\Lambda}}^{(\tau)}(K,h_{_{\Lambda}})$ depends on
the lattice region $\Lambda$, the dimensionless
coupling constant $K$ and the external magnetic field configuration
$h_{_{\Lambda}}$.

\section{The finite system with a film geometry}
\label{film}

Hereafter  we consider only boundary conditions
$\tau=\{\tau_1,p,\cdots,p\}$, with $\tau_1= D, N, ND$ for,
respectively, Dirichlet-Dirichlet, Neumann-Neumann and
Neumann-Diriclet boundary conditions, when
$\vec{k}_{0}=\{1,L_{2},\cdots,L_{d}\}$. By taking the limit
$L_{2},\cdots,L_{d} \rightarrow \infty$ in expression
(\ref{energy}) at fixed $L_{1}=L$ we obtain
\begin{equation}
U_{_{L,d}}^{(\tau_1)}(\phi) := \lim_{L_{2},\cdots, L_{d}
\rightarrow \infty} U_{_{\Lambda}}^{(\tau_1,p,\cdots, p)}(\phi)
\label{energyfilm}
\end{equation}
Next we confine ourselves to the consideration of uniform
magnetic fields, $h(\vec{r})= h$, $\vec{r}\in \Lambda$. By taking
the limit of a film geometry in (\ref{fieldterm}), we obtain
\begin{eqnarray}
 P_{_{L}}^{(\tau)}(K,h;\phi) := \lim_{L_{2},\cdots,
L_{d} \rightarrow \infty}
P_{_{\Lambda}}^{(\tau,p,p)}(K,h_{_{\Lambda}};\phi) \nonumber \\
= {1\over  K L}\sum_{k=1}^{L}{[\hat{h}^{(\tau)}(k)]^{2}
\over \phi + 2\cos\varphi_{_{L}}^{(\tau)}(1) -
2\cos\varphi_{_{L}}^{(\tau)}(k)}, \label{fieldtermF}
\end{eqnarray}
where
\begin{equation}
\hat{h}^{(\tau)}(k):=h\sum_{r=1}^{L}
u_{_{L}}^{(\tau)}(r,k), \quad \tau \in \{D,N, ND \}.
\label{h}
\end{equation}
From Eqs. (\ref{eigenvecD})-(\ref{eigenv}) we obtain explicitly
\begin{equation}
\label{hD}
 \hat{h}^{(D)}(k)=\left\{
\begin{array}{ll}
h \sqrt{\frac{2}{L+1}}\cot\left[\frac{\pi
k}{2(L+1)}\right], & k  \ \ {\mbox{odd}}\\
0, & k \ \ {\mbox{even}},
\end{array}\right.
\end{equation}
\begin{equation}
\label{hN}
\hat{h}^{(N)}(k)=\left\{
\begin{array}{ll}
hL^{1/2}, & k=1\\
-h \sqrt{\frac{2}L}\cos\left[\frac{\pi (k-1)}{2L}\right], &
k=2,\cdots,L,
\end{array}\right.
\end{equation}
and
\begin{equation}
\label{hND} \hat{h}^{(ND)}(k)=
\frac{h}{\sqrt{2L+1}}(-1)^{k-1}\cot\left[{\pi (2k-1)\over
2(2L+1)}\right].
\end{equation}

We just mention that for periodic boundary conditions
the well-known result is $P_{{L}}^{(p)}(K,h;\phi)=h^2/K\phi$.

Note that due to the field dependence of the solution of the mean
spherical constraint for the spherical field $\phi_{L}(K,h)$, one
has to distinguish between two kinds of susceptibilities. The
``fluctuation part" of the susceptibility
\begin{eqnarray}
k_BT\: \chi_{L,\rm fluct}^{(\tau)} (K,h) :&=&\left. -\frac{\partial^2}
{\partial h^2}[\beta f_L^{(\tau)}(K,h)]\right|_{\phi_L^{(\tau)}=const}
\nonumber \\ &=& \left. {1\over 2}\frac{\partial^2}{\partial h^2}
P_L^{(\tau)}(K,h;\phi)]\right|_{\phi=\phi_L^{(\tau)}(K,h)}
\label{flucSUSC}
\end{eqnarray}
measures the fluctuations of the magnetization in the limit of layer geometry
(more precisely, the variance of a properly normalized block-spin in the
corresponding limit Gibbs state) and satisfies the fluctuation-dissipation
theorem. On the other hand, by differentiating twice the
free energy density with respect to the magnetic field,
taking into account the implicit dependence on $h$ through the solution
$\phi_L^{(\tau)}(K,h)$ of the mean spherical constraint,
one obtains the {\it total} magnetic susceptibility per spin,
\begin{eqnarray}
k_BT\: \chi_{L,\rm tot}^{(\tau)}(K,h) := -\frac{{\mathrm d}^2}
{{\mathrm d} h^2}[\beta f_L^{(\tau)}(K,h)]
= k_B T\: \chi_{L,\rm fluct}^{(\tau)} (K,h) \nonumber \\ -
\left. {1\over 2}{\partial^2 \over \partial \phi \partial h}
P_L^{(\tau)}(K,h;\phi)]\right|_{\phi=\phi_L^{(\tau)}(K,h)}
\frac{{\mathrm d} \phi_L^{(\tau)}(K,h)}{{\mathrm d} h}.
\label{MSMSUSC}
\end{eqnarray}
These two susceptibilities coincide in the zero-field case when
there is no spontaneous magnetization in the system.

For the system with a film geometry the mean spherical constraint
(\ref{constraint}) takes the form
\begin{equation}
W_{_{L,d}}^{(\tau)}(\phi) - {\partial \over \partial \phi}
P_{_{L}}^{(\tau)}(K,h;\phi) = K, \label{constraintFB}
\end{equation}
where
\begin{equation}
W_{_{L,d}}^{(\tau)}(\phi) := {1\over L}\sum_{k=1}^{L}W_{d-1}[\phi
+ 2\cos\varphi_{_{L}}^{(\tau)}(1)
-2\cos\varphi_{_{L}}^{(\tau)}(k)] \label{Watson}
\end{equation}
and
\begin{equation}
W_{d-1}(z) = (2\pi)^{-(d-1)} \int_{0}^{2\pi} d
\theta_{1}\int_{0}^{2\pi} \cdots d \theta_{d-1}\: \left[
z+2\sum_{\nu =1}^{d-1} (1-\cos\theta_{\nu})\right]^{-1}.
\label{WatsonRec}
\end{equation}

After the evaluation of $W_{_{L,d}}^{(\tau)}(\phi)$, the
corresponding interaction term
$U_{_{L,d}}^{(\tau)}(\phi_{_{L}}^{(\tau)})$ in the singular (in
the limit $L \rightarrow \infty$) part of the free energy density,
see (\ref{FE}),
\begin{equation}
\beta f_{_{L,\rm sing}}^{(\tau)}(K,h) = \frac12\left \{
U_{_{L,d}}^{(\tau)}(\phi_{_{L}}^{(\tau)}) -
P_{_{L}}^{(\tau)}(K,h;\phi_{_{L}}^{(\tau)}) -
K\phi_{_{L}}^{(\tau)}\right\}, \label{FEInt}
\end{equation}
can be obtained by integration:
\begin{equation}
U_{_{L,d}}^{(\tau)}(\phi_{_{L}}^{(\tau)}) = U_{_{L,d}}^{(\tau)}(\phi_{0}) +
\int_{\phi_{0}}^{\phi_{_{L}}^{(\tau)}} {\mathrm d} \phi
W_{_{L,d}}^{(\tau)}(\phi).
\label{UInt}
\end{equation}
Here $\phi_{_{L}}^{(\tau)} = \phi_{_{L}}^{(\tau)}(K, h)$ is the
solution of equation (\ref{constraintFB}), and $\phi_{0}\geq 0$ is
a suitably chosen constant.

Equations (\ref{fieldtermF}) - (\ref{UInt}) provide the
starting expressions for our further finite-size scaling analysis.

\section{Critical behavior of a three-dimensional film}
\label{finite_size}

At $d=3$ and nonperiodic boundary conditions $\tau = D,N,ND$ at the
surfaces of the film, the interaction term (\ref{Watson}) in the
mean spherical constraint (\ref{constraintFB}) takes the form
\begin{equation}
W_{_{L,3}}^{(\tau)}(\phi) := {1\over L}\sum_{k=1}^{L}W_{2}[
\phi + 2\cos\varphi_{_{L}}^{(\tau)}(1)
-2\cos\varphi_{_{L}}^{(\tau)}(k)].
\label{WatsonLd3}
\end{equation}
This term has been evaluated by using an improved version
\cite{DBA97a} of the method developed by Barber and Fisher
\cite{BF73}, \cite{FB72b}. Following \cite{BF73} we set
\begin{equation}
W_{2}(z):= -(1/4 \pi)\ln{z} + (5/4 \pi)\ln{2} + Q_{2}(z),
\label{DBA13.14}
\end{equation}
where $Q_{2}(z)$ is defined by the above equation. The asymptotic behavior
of $Q_{2}(z)$ as $z \rightarrow 0$ follows
from the well-known one of the Watson integral $W_{2}(z)$:
\begin{equation}
Q_{2}(z)= -{1\over 32\pi}z\ln z + O(z) .
\label{QQQ}
\end{equation}
Now expression (\ref{WatsonLd3}) can be identically rewritten as
\begin {equation}
W_{L,3}^{(\tau)} (\phi) = g_{1}^{(\tau)}(\phi)+g_{2}^{(\tau)}(\phi)+
(5/4\pi)\ln 2,
\label{Watsg1g2tau}
\end{equation}
where
\begin {equation}
g_{1}^{(\tau)}(\phi)= -{1\over {4\pi L}} \sum_{k=1}^{L}
\ln \left(\phi + 2\cos\varphi_{_{L}}^{(\tau)}(1)
-2\cos\varphi_{_{L}}^{(\tau)}(k)\right)
\label{g1tau}
\end{equation}
and
\begin {equation}
g_{2}^{(\tau)}(\phi)= {1\over L} \sum_{k=1}^{L}
Q_{2}\left(\phi + 2\cos\varphi_{_{L}}^{(\tau)}(1) -
2\cos\varphi_{_{L}}^{(\tau)}(k)\right).
\label{g2tau}
\end{equation}

We remark that for the boundary conditions under consideration, the function
(\ref{g1tau}) can be calculated {\it exactly}, and the function (\ref{g2tau})
can be readily evaluated with the aid of the Poisson summation formula.
Here we present the final results.

\subsection{Dirichlet-Dirichlet boundary conditions}
\label{Dirichletbc}

We are interested in the critical regime when $\phi \rightarrow 0^+$ and
$L \rightarrow \infty$, so that $\phi /2 + \cos\varphi_{_{L}}^{(D)}(1) <1$.
Then we set
\begin{equation}
x = \arccos\left[{\phi \over 2} +\cos\left({\pi \over L+1}\right)\right]
\cong \left[{\pi^2 \over (L+1)^2}- \phi \right]^{1/2}.
\label{xD}
\end{equation}
Under the above substitution, the function $g_1^{(D)}(\phi)$, defined
by Eq. (\ref{g1tau}) for $\tau = D$, reads
\begin{equation}
g_1^{(D)}(\phi)= -{1\over 4\pi L} \sum_{k=1}^{L} \ln
\left[2\cos x - 2\cos \left({k\pi \over L+1}\right)\right].
\label{g1sum}
\end{equation}
The sum in the right-hand side can be calculated exactly by making use of
the identity, see \cite{GR73},
\begin{equation}
\cos (nx) - \cos (ny) =(\cos x -\cos y)\prod_{k=1}^{n-1} [2 \cos x
- 2\cos (y+2k\pi /n)],
 \label{g1Da}
\end{equation}
setting here $y=0$, $n=2(L+1)$, and making simple transformations
of the product with the use of the periodicity of the cosine.
Thus one obtains
\begin{equation}
g_1^{(D)}(\phi) =-{1\over 4\pi L}\ln \left[{\sin (L+1)x\over \sin x}\right].
\label{g1D}
\end{equation}

Now we pass to the evaluation of the function $g_2^{(D)}(\phi)$, defined
by Eq. (\ref{g2tau}) at $\tau = D$. Under the substitution (\ref{xD}) it
explicitly reads
\begin{eqnarray}
g_{2}^{(D)}(\phi)={1\over L} \sum_{k=1}^{L}
Q_{2}\left(2\cos x -2\cos{\pi k\over L+1 }\right) \nonumber \\
= {1\over 2L} \sum_{k=1}^{2L+1}
Q_{2}\left(2\cos x -2\cos{\pi k\over L+1 }\right)
-{1\over 2L}Q_{2}(2\cos x + 2).
\label{g2Dsum}
\end{eqnarray}
In deriving the second equality we have used the periodicity of the cosine.
For $\phi < \pi^2/(L+1)^2 \rightarrow 0^+$ as $L\rightarrow \infty$ the last
term obviously yields $Q_2(4)/2L + O(L^{-2})$. By applying the Poisson
summation formula to the sum in the right-hand side of the last equality in
Eq. (\ref{g2Dsum}), changing the integration variable and using the
periodicity of the integrand, we obtain
\begin{eqnarray}
g_{2}^{(D)}(\phi)= {L+1\over L\pi}\int_{\pi/(L+1)}^{\pi}{\mathrm d}\theta \,
Q_2 (2\cos x -2\cos \theta) \nonumber \\
- {1\over 2L} [Q_2(4)- Q_2(\phi)]+\Delta g_{2}^{(D)}(\phi)+ O(L^{-2}),
\label{g2Db}
\end{eqnarray}
where
\begin{eqnarray}
\Delta g_{2}^{(D)}(\phi)=2{L+1\over L\pi}\sum_{q=1}^{\infty}
\int_{\pi/(L+1)}^{\pi}{\mathrm d} \theta \cos[2q(L+1)\theta]
Q_{2}(2\cos x -2\cos \theta) .
\label{g2Dc}
\end{eqnarray}

Consider first the integral in the right-hand side of
Eq. (\ref{g2Db}). Its lower limit cannot be extended to $0$, since we
consider the regime when
\begin{equation}
2\cos x = \phi + 2\cos[\pi /(L+1)] \cong 2 - x^2 < 2,
\label{cosx}
\end{equation}
and the function $Q_2$ is not defined for negative arguments, see Eq.
(\ref{DBA13.14}). That is why we approximate the integrand by
\begin{eqnarray}
Q_{2}(2\cos x -2\cos \theta)\cong Q_{2}(\phi +2-2\cos \theta)
\nonumber \\ -
\left[{\pi\over L+1}\right]^2 Q_2^{\prime}(\phi +2-2\cos \theta),
\label{aprint}
\end{eqnarray}
and notice that the resulting integrals converge when the lower limit
tends to zero. Then, by expressing $Q_2$ in terms of $W_2$ through the
definition (\ref{DBA13.14}), and taking the integral
\begin{eqnarray}
{1\over 4\pi^2}\int_{0}^{\pi}{\mathrm d}\theta \,
\ln (\phi +2 -2\cos \theta)= {1\over 4\pi}
\ln \left\{1+\phi/2 +[\phi(1+ \phi/4)]^{1/2}\right\}\nonumber \\
 = {\phi^{1/2}\over 4\pi } +O(\phi).
\label{g2Dbd}
\end{eqnarray}
we obtain
\begin{equation}
{1\over \pi}\int_{\pi/(L+1)}^{\pi}{\mathrm d}\theta \,
Q_2 (2\cos x -2\cos \theta) \cong W_3(\phi) - {5\ln 2\over 4\pi}
+ {\phi^{1/2}\over 4\pi} + O(L^{-2}).
\label{g2Dbe}
\end{equation}

The integral in the right-hand side of Eq. (\ref{g2Dc}) can be evaluated
by twofold integration by parts. Since the resulting
three terms are $\propto q^{-2}$, the sum over $q$ in Eq. (\ref{g2Dc})
converges and we conclude that
\begin{equation}
\Delta g_{2}^{(D)}(\phi)= O(L^{-2}).
\label{g2Dc2}
\end{equation}

By combining Eqs. (\ref{g2Db}), (\ref{g2Dbe}), (\ref{g2Dc2}), and
taking into account the small argument asymptotic form of $W_3$ and $Q_2$
we obtain
\begin{equation}
g_2^{(D)}(\phi) = K_{c,3} -{5\ln 2 \over 4\pi} +{1\over L}\left[K_{c,3}
-{1\over 2}W_{2}(4)- {7\ln 2\over 8\pi }\right]+O(L^{-2}).
\label{g2Dfin}
\end{equation}

Finally, by substitution of Eqs. (\ref{g1D}) and (\ref{g2Dfin}) into Eq.
(\ref{Watsg1g2tau}), and taking into account Eq. (\ref{xD}), we
obtain the zero-field mean spherical constraint in the form:
\begin{eqnarray}
\ln \left[{\sin (\pi^2 - L^2\phi)^{1/2}\over
(\pi^2 - L^2\phi)^{1/2}}\right] + \ln L= 4\pi L(K_{c,3}-K) \nonumber \\
+ 4\pi \left[K_{c,3}-{1\over 2}W_{2}(4)- {7\ln 2\over 8\pi}\right]
+O(L^{-1}) .
\label{mscD}
\end{eqnarray}
This equation coincides (up to a factor of 2, due to the different choice
of the coupling constant in the Hamiltonian) with the analytical
continuation of Eq. (7.12) in \cite{BF73}, or Eq. (22) in \cite{B74}, from
the domain $L^2\phi >\pi^2$ to
the domain $L^2\phi < \pi^2$. In that work the focus was on the finite-size
scaling behavior and the $\ln L$ term in the left-hand side of
Eq. (\ref{mscD}) was attributed to the finite-size shift of the critical
coupling:
\begin{equation}
K_{m,L}^{(D)}=K_{c,3}-{\ln L \over 4 \pi L}+{1\over L}\left[K_{c,3}
-{1\over 2}W_{2}(4)- {7\ln 2\over 8\pi }\right].
\label{shiftD}
\end{equation}
Then, in terms of the variables $L^2 \phi$ and $L(K_{m,L}^{(D)}-K)$ the
mean spherical constraint (\ref{mscD}) takes the expected finite-size scaling
form (in \cite{BF73} the difference $K_{m,L}^{(D)}-K$ is denoted by
$\Delta{\dot K}$).

However, Chen and Dohm \cite{CD} have noticed that at the bulk
critical temperature, when $K=K_{c,3}$, Eq. (\ref{mscD}) has a
leading-order solution $\phi_L^{(D)}(K_{c,3},0) \sim L^{-3}$. This
behavior follows by assuming $L^2 \phi \rightarrow 0$ as
$L\rightarrow \infty$ and expanding the left-hand side of the mean
spherical constraint up to the leading order:
\begin{equation}
\ln \phi + 3\ln L= 4\pi L(K_{c,3}-K)+ O(1).
\label{mscDlead}
\end{equation}
Note that, if $K_{c,3}>K$, i.e. $L(K_{c,3}-K)\rightarrow\infty$,
the solution of the above equation is
\begin{equation}\label{DbTc}
  \phi \simeq L^{-3}\mbox{e}^{-4\pi L(K_{c,3}-K)+O(1)}.
\end{equation}

To relate the above fact to the critical behavior of the model, we start from
the exact result for the zero-field susceptibility per spin under
Dirichlet-Dirichlet boundary conditions, see Eq. (\ref{PLDh}),
\begin{equation}
\label{hiD}
\chi_L^{(D)}(K,0)=\frac{1}{2J}\left\{\frac{\cot^2(x/2)\sin[(L+1)x]}{L\sin x
[1+\cos(L+1)x]}-\frac{L+1}{2L\sin^2(x/2)} \right\}.
\end{equation}
By expanding the right-hand side for $L^2 \phi \rightarrow 0$ as
$L\rightarrow \infty$, we obtain
\begin{equation}
\label{hiDasympt}
\chi_L^{(D)}(K,0)=\frac{4}{J\pi^2 \phi} + O\left((\phi L)^{-1}\right) +
 O\left(L^{-2}\right).
\end{equation}
Hence, at the {\it bulk} critical temperature one recovers the
leading-order result of \cite{CD}
\begin{equation}
\label{CD} \chi_L^{(D)}(K_{c,3},0) \sim L^3,
\end{equation}
in a full accordance with the predictions of the phenomenological
finite-size scaling, see Eq. (\ref{minusfilm}). For $K_{c,3}>K$
one has $\chi_L^{(D)}(K<K_{c,3},0) \sim L^3\exp[4\pi
L(K_{c,3}-K)]$, i.e. the susceptibility diverges then
exponentially, which agrees with Eq. (\ref{Ybelow}) in the case
$d'=2$.

To understand this anomalous behavior at $T_c$, we recall that the
"standard" finite-size scaling regime takes place when
$L(K_{m,L}^{(D)}-K)=O(1)$, or equivalently, at
\begin{equation}
K = K_{c,3}-{\ln L\over 4\pi L}-{x_1\over L},\quad x_1 =O(1),
\label{KL1D}
\end{equation}
i.e., in a narrow temperature interval of width $O(L^{-1})$ (note
that in the three-dimensional spherical model $\nu=1/(d-2)=1$)
which is shifted by $O(\ln L/L)$ {\it above} the bulk critical
temperature $T_{c,3}$. Thus, the change to the behavior $\chi_L
\sim L^3$, which takes place at the bulk critical point, occurs in
a temperature region of the same width shifted by $O(\ln L/ L)$
{\it below} the finite-size scaling one.

   We emphasize that the result $\chi_L^{(D)}(K_{c,3},0) \sim L^3$  is
characteristic of the layer geometry with {\it two} free surfaces. To show
that we consider below the cases of Neumann-Dirichlet and Neumann-Neumann
boundary conditions.

\subsection{Neumann-Dirichlet boundary conditions}
\label{Neumann_Dirichletbc}

In the critical regime when $\phi \rightarrow 0^+$ and
$L \rightarrow \infty$, so that $\phi/2 +\cos\varphi_{_{L}}^{(ND)}(1)<1$,
we set
\begin{equation}
x = \arccos\left[{\phi \over 2} +\cos\left({\pi \over 2L+1}\right)\right]
\cong \left[{\pi^2 \over (2L+1)^2}- \phi \right]^{1/2}.
\label{xND}
\end{equation}

In this case the exact expression for the function $g_1^{(ND)}(\phi)$,
defined by Eq. (\ref{g1tau}), follows from the identity (4.20) in
\cite{DBA97a}:
\begin{equation}
g_1^{(ND)}(\phi)=-{1\over 4\pi L}\ln \left[{\cos(L+1/2)x\over\cos(x/2)}
\right],
\label{g1ND}
\end{equation}
where $x$ is given by Eq. (\ref{xND}). The evaluation of $g_2^{(ND)}(\phi)$
goes along the same lines as in the previous case with the result
\begin{equation}
g_2^{(ND)}(\phi) = K_{c,3} -{5\ln 2 \over 4\pi} +{1\over 2L}\left[K_{c,3}
- W_{2}(4)- {\ln 2\over 2\pi}\right]+O(L^{-2}).
\label{g2ND}
\end{equation}
Now the zero-field mean spherical constraint takes the form (see Eq. (4.22)
in \cite{DBA97a} at zero surface fields and replace there $2K \rightarrow K$)
\begin{eqnarray}
\ln \left[\cos (\pi^2/4 - L^2\phi)^{1/2}\right] = 4\pi L(K_{c,3}-K)+ 2\pi
\left[K_{c,3}- W_{2}(4)- {\ln 2\over 2\pi}\right]\nonumber \\
 +O(L^{-1}\ln L) .
\label{mscND}
\end{eqnarray}

Comparing the above equation with the analogous Eq. (\ref{mscD}), we see
that now there is no $\ln L/L$ finite-size shift of the critical temperature:
\begin{equation}
K_{m,L}^{(ND)}=K_{c,3}+{1\over 2L}\left[K_{c,3}
-W_{2}(4)- {\ln 2\over 2\pi }\right].
\label{shiftND}
\end{equation}
The mean spherical constraint (\ref{mscND}) takes the expected finite-size
scaling form in terms of the variables $L^2\phi$ and $L(K_{m,L}^{(ND)}-K)$.
Even in the regime $L^2 \phi \rightarrow 0$ as $L\rightarrow \infty$,
by expanding its left-hand side one obtains in the leading order
\begin{equation}
\ln \phi + 2\ln L= 4\pi L(K_{c,3}-K)+ O(1).
\label{mscNDlead}
\end{equation}
Therefore, at the bulk critical temperature one has the standard
finite-size behavior $\phi_L^{(ND)}(K_{c,3},0) \sim L^{-2}$.
Note that below $T_c$, when $L(K_{c,3}-K)\rightarrow\infty$, the solution
of the spherical field equations is
\begin{equation}\label{NbTc}
  \phi \simeq L^{-2}\mbox{e}^{-4\pi L(K_{c,3}-K)+O(1)},
\end{equation}
i.e. it is similar (only the power of $L$ in front of the
exponential term differs) to that of Dirichlet-Dirichlet boundary
conditions, see Eq. (\ref{DbTc}).

The critical behavior of the zero-field susceptibility per spin
under Neumann-Dirichlet boundary conditions follows from the exact
result, see Eq. (\ref{PLNDh}),
\begin{eqnarray}\label{hiND}
\chi_{{L}}^{(ND)}(K,0)=\frac{1}{J}\left\{ 4\cot^2(x/2)\left[
{\tan [(L+1/2)x] \over L\sin x}\right. \right.    \nonumber \\
\left. \left. -\frac{1}{2L(2L+1)\cos^2(x/2)} \right]-
\frac{L+1}{2(2L+1)\sin^2(x/2)}\right\},
\end{eqnarray}
where $x$ is given by Eq. (\ref{xND}). By expanding the right-hand side of
the above equation for $x \ll 1$, we obtain to the leading order
\begin{equation}
\chi_{{L}}^{(ND)}(K,0)\cong \frac{L^2}{J(\pi^2/4 - L^2\phi)}\left[
{\tan (\pi^2/4 - L^2\phi)^{1/2}\over (\pi^2/4 - L^2\phi)^{1/2}}-1\right].
\label{hiNDas}
\end{equation}
        From Eq. (\ref{mscND}) it is clear that for any finite value of
$L(K_{c,3} -K)$ the susceptibility diverges as $L^2$ when
$L\rightarrow \infty$. Note that $(\pi^2/4 -
L^2\phi)^{1/2}\rightarrow 0^+$ only when
$L(K_{m,L}^{(ND)}-K)\rightarrow 0$. Therefore, at the shifted
critical temperature one obtains the simple result
\begin{equation}
\chi_{{L}}^{(ND)}(K_{m,L}^{(ND)},0)\cong \frac{L^2}{3J}.
\label{hiNDshift}
\end{equation}

Thus we conclude that in the presence of only one free (Dirichlet) surface
the mean spherical model has the usual
finite-size critical behavior. It is instructive to see whether
this behavior will change in the presence of two equivalent
surfaces with Neumann boundary conditions.

\subsection{Neumann-Neumann boundary conditions}
\label{Neumann_Neumannbc}

The interaction term (\ref{WatsonLd3}) in this case has been treated
completely and rigorously in \cite{DBA97a}. Since
$\cos\varphi_{_{L}}^{(N)}(1)=1$, see (\ref{eigenv}), we set
\begin{equation}
x = \cosh^{-1}(1+\phi /2) \cong  \phi^{1/2},
\label{xN}
\end{equation}
where $\cosh^{-1}$ denotes the inverse function of $\cosh$. The exact
expression for the function $g_{1}^{(N)}(\phi)$ is given by Eq. (3.13) of
\cite{DBA97a}. It reads
\begin {equation}
g_{1}^{(N)}(\phi)= -{1\over 4\pi L}\left\{\ln \phi+\ln \left[{\sinh Lx \over
\sinh x}\right]\right\} .
\label{g1N}
\end{equation}

In the critical regime $\phi \rightarrow 0$ as $L \rightarrow \infty$ the
result for the function $g_{2}^{(N)}(\phi)$ is (see Eq. (3.18) in
\cite{DBA97a} under the replacement $2K_c \rightarrow K_c$):
\begin{equation}
g_{2}^{(N)}(\phi)=K_{c}-{5\ln 2 \over 4\pi} -{1\over 2L}
\left[W_{2}(4)-{3\ln 2\over 4\pi}\right] + O(L^{-2}).
\label{g2N}
\end{equation}

Thus, by substitution of equations (\ref{g1N}) and (\ref{g2N}) into
(\ref{Watsg1g2tau}), in the limit $\phi \rightarrow 0$, $L \rightarrow
\infty$, we obtain in the leading order the following finite-size scaling form
of the zero-field mean spherical constraint:
\begin{equation}
\ln \left[L\phi^{1/2} \sinh (L\phi^{1/2})\right]= 4\pi L(K_{m,L}^{(N)} - K),
\label{mscN}
\end{equation}
where $K_{m,L}^{(N)}$ is the shifted critical coupling,
\begin{equation}
K_{m,L}^{(N)}= K_{c,3} + {\ln L\over 4\pi L} -{1\over 2 L}
\left[W_{2}(4)-  {3 \ln 2\over 2}\right].
\label{shiftN}
\end{equation}

Comparing Eqs. (\ref{shiftN}) and (\ref{shiftD}) we see that under
Neumann-Neumann boundary conditions the finite-size scaling region of width
$O(L^{-1})$ is shifted {\it below} the bulk critical temperature $T_{c,3}$
by $O(\ln L/ L)$. Since at the bulk critical point, when $K=K_{c,3}$,
the right-hand side of Eq. (\ref{mscN}) goes to plus infinity like $\ln L$ as
$L\rightarrow \infty$, we conclude that $L\phi^{1/2}\rightarrow \infty$,
although $\phi\rightarrow 0^+$. In this regime Eq. (\ref{mscN}) simplifies to
\begin{equation}
\ln \phi^{1/2} + L\phi^{1/2}= O(1),
\label{mscNas}
\end{equation}
which yields
\begin{equation}
\phi_L^{(N)}(K_{c,3}) \sim  (\ln L /L)^2 .
\label{phiNas}
\end{equation}
This, at the bulk critical temperature, the finite-size critical
behavior of the spherical field under Neumann-Neumann boundary
conditions becomes logarithmically modified. Note that below
$T_c$, i.e. when $L(K_{c,3}-K)\rightarrow\infty$, the solution of
the spherical field equation (\ref{mscN}) is
\begin{equation}\label{NNbTc}
  \phi \simeq L^{-1}\mbox{e}^{-4\pi L(K_{c,3}-K)+O(1)},
\end{equation}
i.e. the leading-order behavior of $\phi$ again differs only in
the power of $L$ in front of the exponential term, compare with
Eqs. (\ref{DbTc}) and (\ref{NbTc}).

The corresponding critical behavior of the zero-field susceptibility follows
from the exact expression, see (\ref{PLNh}),
\begin{equation}\label{hiN}
\chi_{{L}}^{(N)}(K,0)= \frac{1}{J}\left\{\frac{1}{\phi}+
\frac{1+\cosh x}{L^2 \sinh x}\left[L\coth(Lx)-\coth x\right]-
\frac{L-1}{L^2}\right\},
\end{equation}
where $x$ is given by Eq. (\ref{xN}). By expanding this expression for
$x\cong \phi^{1/2}\rightarrow 0^+$ and $Lx\cong L\phi^{1/2}\rightarrow
\infty$, we obtain the following modified leading-order behavior of the
finite-size zero-field susceptibility at the bulk critical point:
\begin{equation}\label{hiNas}
\chi_{{L}}^{(N)}(K_{c,3},0)\cong
\frac{1}{J\phi_L^{(N)}(K_{c,3})}\cong {L^2\over J(\ln L)^2},
\end{equation}
in a full accordance with Eq. (\ref{plussm}).

 We emphasize that the above behavior takes place at the
bulk critical temperature, in a temperature region of width
$O(L^{-1})$ shifted by $O(\ln L/ L)$ {\it above} the finite-size
scaling one. It will be further modified, although not radically,
if one considers the temperature interval located at $p>0$ times
the same shift above $T_{c,3}$. Then one obtains from Eq. (\ref{mscN}),
up to the leading order,
\begin{equation}
\phi_L^{(N)}(K_{L,p}) \sim  [(1+p)\ln L /L]^2 ,
\label{phiNasp}
\end{equation}
and hence the corresponding reduction of the finite-size zero-field
susceptibility (\ref{hiNas}) by a factor of $(1+p)^{-2}$.

\section{Discussion}

In the current article we have presented the predictions of the
phenomenological finite-size scaling for systems with
asymptotically large shift of the bulk critical temperature. We
have shown that in such systems the behavior of the zero-field
susceptibility at the bulk critical point depends crucially on the
sign of the shift - the positive shift leads to the reduction of
the "standard" divergence of $\chi_L^{(\tau)}\sim L^{\gamma/\nu}$,
while the negative shift leads to a stronger divergence. We
have verified our considerations on the example of the
three-dimensional spherical model under Dirichlet-Dirichlet,
Dirichlet-Neumann, and Neumann-Neumann boundary conditions. A
Dirichlet surface leads to a negative shift of the critical coupling,
$-\ln L/(8\pi L)$, and a Neumann surface - to a positive one,
$\ln L/(8\pi L)$ \cite{D93}. So, for a film geometry under Dirichlet-Dirichlet
boundary condition the shift $\varepsilon_L^{(D)}=-\ln L/(4\pi L)$
leads to $\chi_L^{(D)}(K_{c,3},0) \sim L^3$, (see Eq. \ref{CD}),
in full accordance with our phenomenological predictions (see
Eqs. (\ref{Ybelow}) and (\ref{minusfilm})). For Neumann-Neumann
boundary conditions the shift is positive, i.e.
$\varepsilon_L^{(N)}=\ln L/(4\pi L)$, wherefrom
$\chi_{{L}}^{(N)}(T_c)\cong L^2/[ J(\ln L)^2]$, in a full
agreement with the phenomenoligal prediction given by Eq.
(\ref{plussm}).

We emphasize that, in order to make concrete phenomenological
prediction for the finite-size behavior of the zero-field
susceptibility in systems with asymptotically large shift of the
critical temperature, use has been made of the the size dependence
of $\chi_L^{(p)}$ for $T<T_c$. We have supposed that the leading
behavior of $\chi_L^{(\tau)}$ will be the same under other
boundary conditions. We have verified this assumption on the
example of the spherical model. It is highly desirable to have the
corresponding results for other models too.

One might ask why is the shift of the critical coupling in the
spherical model positive under Neumann-Neumann and negative under
Dirichlet-Dirichlet boundary conditions. Indeed, this contradicts
the general expectations based on arguments like that the missing
neighbors in a ferromagnetic system should  reduce its critical
temperature. In order to understand the above facts, let us note
that the observed behavior is in agreement with the length of the
spins near the boundary. This length is reduced near a Dirichlet
boundary and enlarged near a Neumann one \cite{DBA97b} (there
$<\sigma^2>\simeq 1.34$ \cite{DBA97b}; similar estimation for the
Dirichlet boundary gives $<\sigma^2>\simeq 0.83$). Then, since the
total length of all the spins is fixed, that leads to spins in the
main part of the system being larger than $1$ under Dirichlet and
smaller than $1$ under Neumann boundary conditions. As a result,
an effective interaction is taking place with spins which length
is not equal to $1$. In turn, this produces a shift in the
"critical temperature" of the finite system which is positive for
Dirichlet and negative for Neumann boundary conditions, contrary
to what one would expect for a system with a fixed length of the
spins.

Finally, let us recall that the infinite translational invariant
spherical model is equivalent to the $n\rightarrow \infty $ limit
of the corresponding
system of $n$%
-component vectors \cite{S68},\cite{KT71}, but the spherical model
with surfaces (or, more generally, without translation-invariant
symmetry) is in fact {\em not} such a limit \cite{K73}. In other
words, the spherical model under nonperiodic boundary conditions
is not in the same surface universality class as the corresponding
$O(n)$ model in the limit $n\rightarrow \infty $, in contrast with
the bulk universality classes. The last becomes apparent when one
investigates surface phase transitions for an $O(n)$ model in the
limit $n\rightarrow \infty $. In that case in the limit
$n\rightarrow\infty$ one obtains \cite{Bi83} $\Delta _1^{\rm
o}=1/(d-2)$ (i.e. $\Delta _1^{\rm o}=1$ for $d=3$) for ordinary
and $\Delta _1^{\rm sb}=2/(d-2)$ (i.e. $\Delta _1^{\rm sb}=2$ for
$d=3$) for special phase transitions, while $\Delta _1^{\rm
o}=1/2$ and $\Delta _1^{\rm sb}=3/2$ \cite{DBA97a}, \cite{DBA97b}
for the three-dimensional spherical model. It is believed that the
corresponding equivalence will be recovered if one imposes
spherical constraints in a way which ensures that the mean square
value of each spin of the system is the same \cite{K73}
(unfortunately such a model is rather untractable analytically).
The introduction of just one additional spherical field that fixes
the mean square values of the spins at the boundaries changes the
surface critical exponents \cite{DBA97b} but is not enough to
recover the correspondence to the $O(n)$ models. Unfortunately,
the finite size scaling properties even of that analytically
tractable model have not been investigated.

\section*{Acknowledgements}

The authors thank Prof. M. E. Fisher for drawing their attention
to Ref. \cite{FB72b} and Prof. S. Dietrich for the fruitful
discussions.

D. Dantchev acknowledges the  hospitality of Max-Planck-Institute
for Metals Research in Stuttgart  as well as the financial support
of the Alexander von Humboldt Foundation.

\appendix
\renewcommand{\theequation}{\Alph{section}$.$\arabic{equation}}
\section{Calculation of the field term}
\label{asums}

\subsection{Dirichlet-Dirichlet boundary conditions}

In uniform magnetic field $h$ and in the regime when
$\cos x := \phi/2+\cos \varphi_L^{(D)}(1) <1$, the field term Eq.
(\ref{fieldtermF}) has the explicit form (assuming $L$ odd)
\begin{equation}
P_{_{L}}^{(D)}(K,h;\phi)= {h^2\over KL(L+1)}\sum_{k=0}^{(L-1)/2}
{\cot^2 {\pi(2k+1)\over 2(L+1)} \over
\cos x - \cos {\pi (2k+1)\over L+1}}.
\label{PLDsum}
\end{equation}
By making use of the elementary identity
\begin{equation}
\cot^2(\alpha/2) = {2\over 1-\cos \alpha} - 1,
\label{elem}
\end{equation}
the above expression can be rewritten as
\begin{equation}
P_{_{L}}^{(D)}(K,h;\phi)= {h^2\over K}\left\{ {2\over 1-\cos x}
\left[S_L^{(D)}(x) - S_L^{(D)}(0)\right] - S_L^{(D)}(x)\right\},
\label{PLDa}
\end{equation}
where
\begin{equation}
S_L^{(D)}(x)= {1\over L(L+1)}\sum_{k=0}^{(L-1)/2}
{1 \over \cos x - \cos {\pi (2k+1)\over L+1}}.
\label{PLDb}
\end{equation}
To calculate the sum $S_L^{(D)}(x)$, we use the identity (\ref{g1Da}) and
set there $n=L+1$ and $y=\pi /(L+1)$. This yields
\begin{equation}
\prod_{k=0}^{L}\left[\cos x - \cos {\pi (2k+1)\over L+1}\right] =
2^{-L}[\cos (L+1)x+1].
\label{PLDc}
\end{equation}
Since the left-hand side of the above identity equals
\begin{equation}
\left\{\prod_{k=0}^{(L-1)/2}\left[\cos x - \cos {\pi (2k+1)\over L+1}
\right]\right\}^2,
\label{PLDd}
\end{equation}
we obtain
\begin{equation}
\prod_{k=0}^{(L-1)/2}\left[\cos x - \cos {\pi (2k+1)\over L+1}\right] =
2^{-L/2}[\cos (L+1)x+1]^{1/2}.
\label{PLDe}
\end{equation}
Next, by taking logarithm of both sides of Eq. (\ref{PLDe}) and
differentiating the result with respect to $x$, we obtain
\begin{equation}
S_L^{(D)}(x)= {\sin (L+1)x \over 2L\sin x[1+\cos (L+1)x]}.
\label{PLDf}
\end{equation}
Hence
\begin{equation}
S_L^{(D)}(0)= \lim_{x\rightarrow 0}S_L^{(D)}(x)= {L+1\over 4L}.
\label{PLDg}
\end{equation}
Finally, by inserting Eqs. (\ref{PLDf}) and (\ref{PLDg}) into Eq.
(\ref{PLDa}), and making elementary transformations based on the
identity (\ref{elem}), we obtain the exact result
\begin{equation}
P_{_{L}}^{(D)}(K,h;\phi)= {h^2\over 2K}
\left\{\frac{\cot^2(x/2)\sin[(L+1)x]}{L\sin x
[1+\cos(L+1)x]}-\frac{L+1}{2L\sin^2(x/2)} \right\}.
\label{PLDh}
\end{equation}

\subsection{Neumann-Dirichlet boundary conditions}

In uniform magnetic field $h$ and in the regime when
$\cos x := \phi/2+\cos \varphi_L^{(ND)}(1) <1$, the field term Eq.
(\ref{fieldtermF}) has the following explicit form
\begin{equation}
P_{_{L}}^{(ND)}(K,h;\phi)= {h^2\over KL(2L+1)}\sum_{k=0}^{L-1}
{\cot^2 {\pi(2k+1)\over 2(2L+1)} \over
\cos x - \cos {\pi (2k+1)\over 2L+1}}.
\label{PLNDsum}
\end{equation}
By making use of the elementary identity (\ref{elem})
the above expression can be rewritten as
\begin{equation}
P_{_{L}}^{(ND)}(K,h;\phi)= {h^2\over K}\left\{ {2\over 1-\cos x}
\left[S_L^{(ND)}(x) - S_L^{(ND)}(0)\right] - S_L^{(ND)}(x)\right\},
\label{PLNDa}
\end{equation}
where
\begin{equation}
S_L^{(ND)}(x)= {1\over 2L(2L+1)}\sum_{k=0}^{L-1}
{1 \over \cos x - \cos {\pi (2k+1)\over 2L+1}}.
\label{PLNDb}
\end{equation}
To calculate the sum $S_L^{(ND)}(x)$, we use the identity (\ref{g1Da}) and
set there $n=2L+1$ and $y=\pi /(2L+1)$. This yields
\begin{equation}
\prod_{k=0}^{2L}\left[\cos x - \cos {\pi (2k+1)\over 2L+1}\right]
= 2^{-2L}[\cos (2L+1)x+1]. \label{PLNDc}
\end{equation}
Since the left-hand side of the above identity equals
\begin{equation}
(\cos x +1)\left\{\prod_{k=0}^{L-1}\left[\cos x - \cos {\pi (2k+1)\over 2L+1}
\right]\right\}^2,
\label{PLNDd}
\end{equation}
we obtain
\begin{equation}
\prod_{k=0}^{L-1}\left[\cos x - \cos {\pi (2k+1)\over 2L+1}\right] =
2^{-L}\left[{\cos (2L+1)x+1\over \cos x +1}\right]^{1/2}.
\label{PLNDe}
\end{equation}
Next, by taking logarithm of both sides of Eq. (\ref{PLNDe}) and
differentiating the result with respect to $x$, we obtain
\begin{equation}
S_L^{(ND)}(x)= {\sin (2L+1)x \over 4L\sin x[1+\cos (2L+1)x]}-
{1\over 4L(2L+1)(1+\cos x)}. \label{PLNDf}
\end{equation}
Hence
\begin{equation}
S_L^{(ND)}(0)= \lim_{x\rightarrow 0}S_L^{(ND)}(x)= {L+1\over
2(2L+1)}. \label{PLNDg}
\end{equation}
Finally, by inserting Eqs. (\ref{PLNDf}) and (\ref{PLNDh}) into
Eq. (\ref{PLNDa}), and making elementary transformations, we obtain the
exact result
\begin{eqnarray}
P_{_{L}}^{(D)}(K,h;\phi)= {h^2\over K}
\left\{4\cot^2(x/2)\left[\frac{\tan[(L+1/2)x]}{L\sin x}\right. \right.
\nonumber \\ \left. \left.
-{1\over 2L(2L+1) cos^2(x/2)}\right] -\frac{L+1}{2(2L+1)\sin^2(x/2)}\right\}.
\label{PLNDh}
\end{eqnarray}

\subsection{Neumann-Neumann boundary conditions}

This case differs from the previous two ones in that now
$\cos \varphi_L^{(N)}(1) = 1$ and we have to set $\cosh x := 1+ \phi/2 >1$.
In uniform magnetic field $h$, the field term Eq. (\ref{fieldtermF}) has the
explicit form
\begin{equation}
P_{_{L}}^{(N)}(K,h;\phi)= {h^2\over K \phi} +{2h^2\over K}S_L^{(N)}(x),
\label{PLNsum}
\end{equation}
where
\begin{equation}
S_L^{(N)}(x)= {1\over L^2}\sum_{k=1}^{L-1}
{\cos^2{\pi k\over 2L} \over \cosh x - \cos (\pi k/ L)}.
\label{PLNa}
\end{equation}
By making use of the elementary identity
\begin{equation}
\cos^2(\alpha/2) = {1\over 2}(1+\cosh x)-{1\over 2}(\cosh x-\cos \alpha),
\label{elemN}
\end{equation}
the above expression can be rewritten as
\begin{equation}
S_L^{(N)}(x)= {1+\cosh x\over 2L^2}\sum_{k=1}^{L-1}
{1 \over \cosh x -\cos (\pi k /L)}- {L-1\over 2L^2}.
\label{PLNb}
\end{equation}
To calculate the above sum, we start from the identity \cite{GR73}
\begin{equation}
\prod_{k=1}^{n-1}\left[z^2 -2z\cos {\pi k\over n} +1\right] =
{z^{2n} -1\over z^2 -1}.
\label{idenN}
\end{equation}
By setting here $z=\exp(x)$, $n=L$, and performing elementary transformations,
we obtain
\begin{equation}
\prod_{k=1}^{L-1}\left[\cosh x -\cos {\pi k\over L}\right] =
2^{-L+1}{\sinh (Lx) \over \sinh x}.
\label{idenN1}
\end{equation}
Next, by taking logarithm of both sides of Eq. (\ref{PLDe}) and
differentiating the result with respect to $x$, we obtain
\begin{equation}
S_L^{(N)}(x)= {1+\cosh x\over 2L^2\sinh x}[L\coth (Lx)-\coth x]-
{L-1\over 2L^2}.
\label{PLNf}
\end{equation}
Finally, by inserting equation (\ref{PLNf})  into equation
(\ref{PLNsum}), we obtain the exact result
\begin{equation}
P_{_{L}}^{(N)}(K,h;\phi)= {h^2\over K}\left\{\frac{1}{\phi}+
\frac{1+\cosh x}{L^2 \sinh x}\left[L\coth(Lx)-\coth x\right]-
\frac{L-1}{L^2}\right\}.
\label{PLNh}
\end{equation}

\end{document}